\documentclass[jkps,preprint,fleqn,showpacs,showkeys]{revtex4}
\usepackage{graphicx}
\usepackage{amssymb}
\usepackage{amsmath}

\begin{document}
\setcounter{page}{0}
\title{Cluster Monte Carlo Study of Magnetic Dipoles}
\author{Seung Ki \surname{Baek}}
\email{garuda@tp.umu.se}
\thanks{Fax: +46-70-786-6797}
\affiliation{Integrated Science Laboratory, Department of Physics, Ume{\aa}
University, 901 87 Ume{\aa}, Sweden}

\date[]{Received 10 May 2011}

\begin{abstract}
We implement a cluster-update Monte Carlo algorithm to simulate magnetic
dipoles of the $XY$-spin type confined in a two-dimensional plane.
The long-range character and anisotropy in the dipole interaction
are handled by using the Luijten-Bl\"ote algorithm and the
Dotsenko-Selke-Talapov algorithm, respectively. We have checked the
performance of this cluster-update algorithm in comparison to the Metropolis
algorithm and found that it equilibrated the system faster in terms of the
number of flipped spins, although the overall computational complexity of
the problem remained the same.
\end{abstract}


\maketitle

\section{INTRODUCTION}
Magnetism has been one of the most important subjects in physics,
and many of technological applications are based upon ordered behaviors of
magnetic materials. This leads us to both a theoretical and practical
question about how to understand collective behaviors observed in magnetic
systems.
A particularly interesting case to us is rare-earth compounds in which magnetic
spins at low temperatures can be basically regarded as two-dimensional
(2D)~\cite{lynn} and the exchange interaction is relatively weak~\cite{mac}
because one can easily make use of theoretical frameworks developed to
study such continuous spin models in statistical physics.
Let us consider a square lattice of $XY$-like
spins governed by the dipole interaction. If the linear size of the
lattice is $L$, the total number of spins will be $N=L^2$, and the
corresponding Hamiltonian is given as follows:
\begin{equation}
H = J \sum_{i \neq j}^N \left[ (\boldsymbol{s}_i \cdot \boldsymbol{s}_j)
r_{ij}^2 - 3 (\boldsymbol{s}_i \cdot \boldsymbol{r}_{ij}) (\boldsymbol{s}_j
\cdot \boldsymbol{r}_{ij}) \right] / r_{ij}^5,
\label{eq:h}
\end{equation}
where $J(>0)$ represents interaction strength
and the summation runs over every
distinct spin pair of $\boldsymbol{s}_i$ and $\boldsymbol{s}_j$ at
$\boldsymbol{r}_i$ and $\boldsymbol{r}_j$, respectively.
The distance between this spin pair is denoted as $r_{ij} = \left|
\boldsymbol{r}_i - \boldsymbol{r}_j \right|$.
Since the periodic boundary condition is employed to reduce unwanted boundary
effects, the relative displacement $\boldsymbol{r}_{ij}$ between
$\boldsymbol{r}_i$ and $\boldsymbol{r}_j$ is chosen as the one with
minimal length among every possible pair of their periodic images. In case
that more than one periodic image of a spin has the same minimal length
from another spin, ambiguity can enter in defining the interaction within
this pair due to the anisotropy manifested in the second term of
Eq.~(\ref{eq:h}). To be simple, we neglect the interaction in such a case,
and this choice does not hurt any essential properties of the
system.
It is notable that the interaction energy between $\boldsymbol{s}_i$ and
$\boldsymbol{s}_j$ has an overall distance dependence as $r_{ij}^{-3}$
and is determined by both their phase difference and the relative
displacement, $\boldsymbol{r}_{ij}$.
In a numerical analysis, the long-range character implies
computational complexity of $O(N^2)$ for the simple Metropolis algorithm
(see, however, Ref.~\cite{saaki} for a possible modification of this
approach). For this reason, it has not been easy to precisely determine the
physical properties of the phase transition in a dipole
system~(see Refs.~\cite{fer,tomita,honey} and references therein), and it
still remains to be investigated from an algorithmic point of view.
 
In this work, we try to implement a
Wolff-type single-cluster update algorithm~\cite{wolff} for a dipole
system to challenge this issue. Specifically,
we combine the Luijten-Bl{\"o}te (LB) algorithm~\cite{lb}
and the Dotsenko-Selke-Talapov (DST) algorithm~\cite{dst}
to analyze Eq.~(\ref{eq:h}) and check the results in comparison to
the Metropolis single-spin update algorithm. This work is
organized as follows: We begin with the LB algorithm in Sec.~\ref{sub:lb}
and the DST algorithms will be given in Sec.~\ref{sub:dst}. Then we combine them
to construct the cluster-update algorithm and present results in
Sec.~\ref{sub:com}. This work is summarized in Sec.~\ref{sec:sum}.

\section{METHOD AND RESULTS}
\label{sec:alg}

\subsection{Luijten-Bl{\"o}te Algorithm}
\label{sub:lb}

Let us begin with a 2D ferromagnetic system described as
\begin{equation}
H = - \sum_{i \neq j} J_{ij} \boldsymbol{s}_i \cdot \boldsymbol{s}_j,
\label{eq:iso}
\end{equation}
where $J_{ij} \equiv J / r_{ij}^3$ and each index runs from 1 to $N$.
We regard $\boldsymbol{s}_n$ as an Ising spin for a while.
If one directly applies the Wolff algorithm, the updating procedure would be
as follows.
\begin{enumerate}
\item Pick a spin randomly and add its index into a stack.
\item \label{itm:get} Retrieve an element $i$ from the stack.
\item \label{itm:chk} For every other spin $\boldsymbol{s}_j$ ($j \neq i$)
in the system, add its index $j$ into the stack with probability $P_{i,j}
= \left[ 1-\exp(-2J_{ij}/k_B T) \right] \delta_{\boldsymbol{s}_i,
\boldsymbol{s}_j}$ with the Boltzmann constant $k_B$.
\item If the stack is not empty, go to Step~\ref{itm:get}. Otherwise, flip
the cluster.
\end{enumerate}
It is obvious that Step~\ref{itm:chk} spends time proportional to $O(N^2)$
for every retrieval if the program checks whether $\boldsymbol{s}_i =
\boldsymbol{s}_j$ for each spin pair. The idea of the LB algorithm is that
we may first assume $\boldsymbol{s}_i = \boldsymbol{s}_j$ to calculate 
$P_{i,j}' = 1-\exp(-2J_{ij}/k_B T)$, which is fixed throughout the whole
computation. Note that the index $i$ is irrelevant because every point is
equivalent under the periodic boundary condition.
Hence, instead of checking this probability for every spin, we
can build a probability table in advance to correctly pick up possible
$\boldsymbol{s}_j$'s in terms of a position relative to $\boldsymbol{s}_i$.
This saves time to $O(N \log N)$, where $\log N$ appears in looking up the
table with the binary search~\cite{fukui}.
If the picked $\boldsymbol{s}_j$
does not point in the same direction as $\boldsymbol{s}_i$, we do not add it
into the stack because we need to recover $P_{i,j} = P_{i,j}' \times
\delta_{\boldsymbol{s}_i, \boldsymbol{s}_j}$.
Specifically, the prescription in Ref.~\cite{lb} can be written as follows:
\begin{enumerate}
\item Assume that we have chosen $\boldsymbol{s}_i$ from which we grow a
cluster.
\item A spin $\boldsymbol{s}_n$ ($j \neq i$) can
be chosen with probability $Q_0(n) = (1-P_{i,1}') \times (1-P_{i,2}') \times
\cdots \times (1-P_{i,n-1}') \times P_{i,n}'$. Let's say $\boldsymbol{s}_j$
is picked up by performing this step.
\item The next spin is then selected according to a new probability
distribution, $Q_j(n) = (1-P_{i,j+1}') \times (1-P_{i,j+2}') \times \cdots
\times (1-P_{i,n-1}') \times P_{i,n}'$. This step is repeated: that is,
whenever a spin $\boldsymbol{s}_l$ is selected, we work with
$Q_l(n)$ to find the next one.
This repetition stops when no more spins are picked up.
\end{enumerate}
By doing this, each spin $\boldsymbol{s}_j$ is chosen with its own correct
probability, $P_{i,j}'$. To demonstrate this, let us set $b_n = 1$ when
$\boldsymbol{s}_n$ is picked up and $b_n = 0$ in the
other case. If we denote $Pr(b_1,b_2,\ldots,b_n)$ as the probability of
each possible event, the sum of all the probabilities to choose
$\boldsymbol{s}_2$ then reads as \[ Pr(0,1) + Pr(1,1) = (1-P_{i,1}') P_{i,2}' +
P_{i,1}' P_{i,2}' = P_{i,2}'.\] Likewise, the total probability to select
$\boldsymbol{s}_3$ is
\begin{eqnarray*}
&&Pr(0,0,1) + Pr(0,1,1) + Pr(1,0,1) + Pr(1,1,1)\\
&=& (1-P_{i,1}')(1-P_{i,2}') P_{i,3}'
+ (1-P_{i,1}') P_{i,2}' P_{i,3}'
+ P_{i,1}' (1 - P_{i,2}') P_{i,3}'
+ P_{i,1}' P_{i,2}' P_{i,3}'\\
&=& P_{i,3}',
\end{eqnarray*}
and one can readily generalize this for any arbitrary $\boldsymbol{s}_j$. It
is also clear that this holds true no matter how one indexes spins as long
as the index is used consistently throughout the calculation even though
Ref.~\cite{lb} makes it with respect to physical distances.

In picking up a spin index from such a procedure, it is convenient to work
with a cumulative distribution, that is,
\begin{eqnarray}
&&C_j(n) = \sum_{l=j+1}^n Q_j(l) = 1 - \exp \left[ -2\sum_{l=j+1}^n
J_{il}/k_B T \right],\nonumber\\
&&-\frac{1}{2} \ln \left[ 1-C_j(n) \right] = \sum_{l=j+1}^n J_{il}/k_B T
\equiv S_j(n).
\label{eq:cj}
\end{eqnarray}
and one will find which spin should be picked by comparing $C_j(n)$ with a
random number uniformly drawn over $[0,1)$. A beauty of the LB algorithm is
that updating of $C_j(n)$ with picking a spin hardly needs any additional
computation because $j$ only enters as a starting index of the summation in
Eq.~(\ref{eq:cj}) so that $S_0(n) - S_0(j) = S_j(n)$ for $1 \le j < n$.
In other words, it suffices to compute $S_0(n)$ once and to memorize it for every
$n$ in advance of the Monte Carlo iterations. One needs to reset the starting
index as $j$ when having picked up spin $j$, and then
one will get the correct $J_{il}/k_B T$ (accordingly, correct
$P_{i,l}'$) for spin $l>j$ each time by using Eq.~(\ref{eq:cj}).
Note that it is most natural to define $S_0(0)$ as
zero. Then, the LB algorithm for solving Eq.~(\ref{eq:iso}) modifies the
Wolff algorithm given above in the following way:
\begin{enumerate}
\item Choose $\boldsymbol{s}_i$ placed at a certain position,
and make an array of partial
sums $S_0(n)$ with relative displacements from this chosen spin site.
\item Pick a spin randomly and add its index into a stack. Set $z$ as zero.
\item \label{itm:get2} Retrieve an element $i$ from the stack and do the
following:
\begin{enumerate}
\item \label{itm:cum} Draw a uniform random number $u \in [0,1)$ and find an
index $w$ that satisfies $S_0(w) \le -\frac{1}{2} \ln ( 1-u ) + S_0(z) <
S_0(w+1)$. If there is no such $w$, terminate this loop for $i$. Otherwise,
set $z$ as $w$ for the next iteration.
\item Since $w$ indicates only a relative position with respect to $i$,
translate it into the actual position $w'$.
\item Add $w'$ into the stack if $\boldsymbol{s}_i = \boldsymbol{s}_{w'}$;
go to step~\ref{itm:cum}.
\end{enumerate}
\item If the stack is not empty, go to step~\ref{itm:get2}. Otherwise, flip
the cluster.
\end{enumerate}
\begin{figure}
\includegraphics[width=0.35\textwidth]{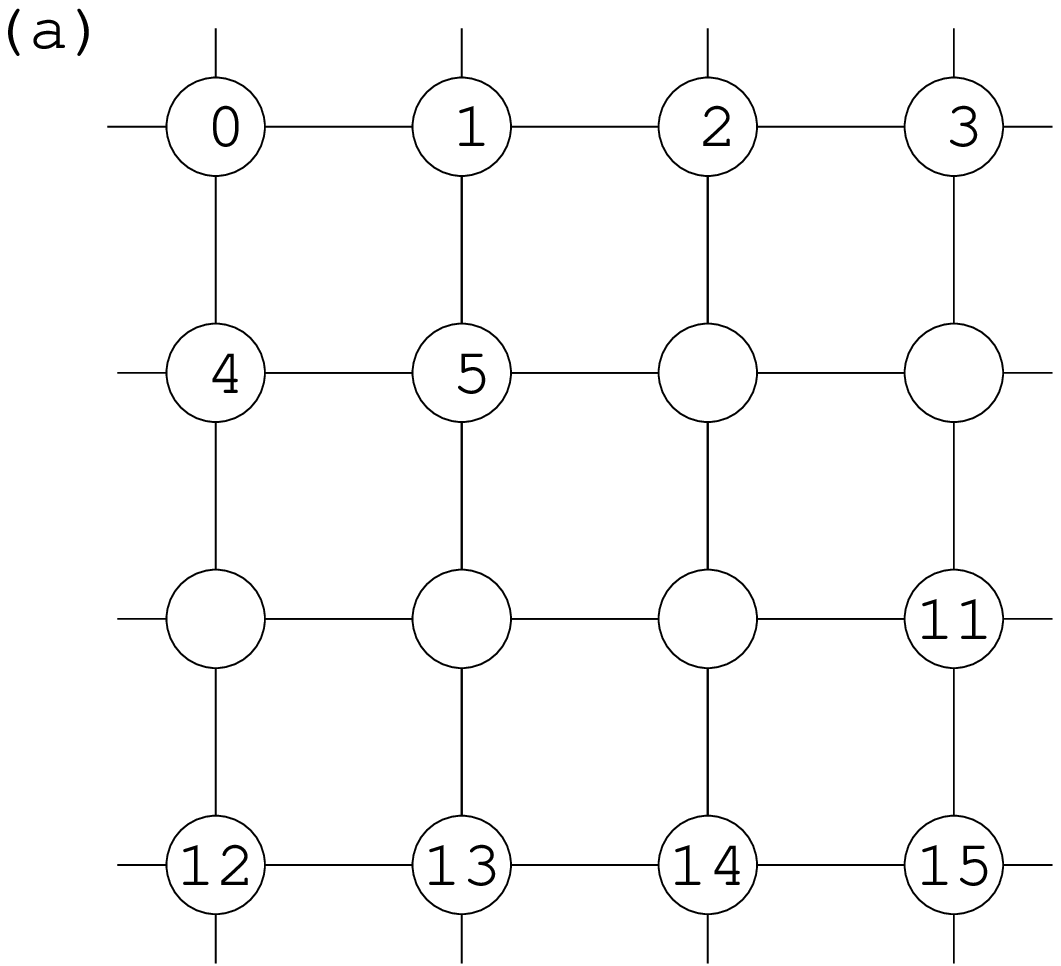}
\includegraphics[width=0.35\textwidth]{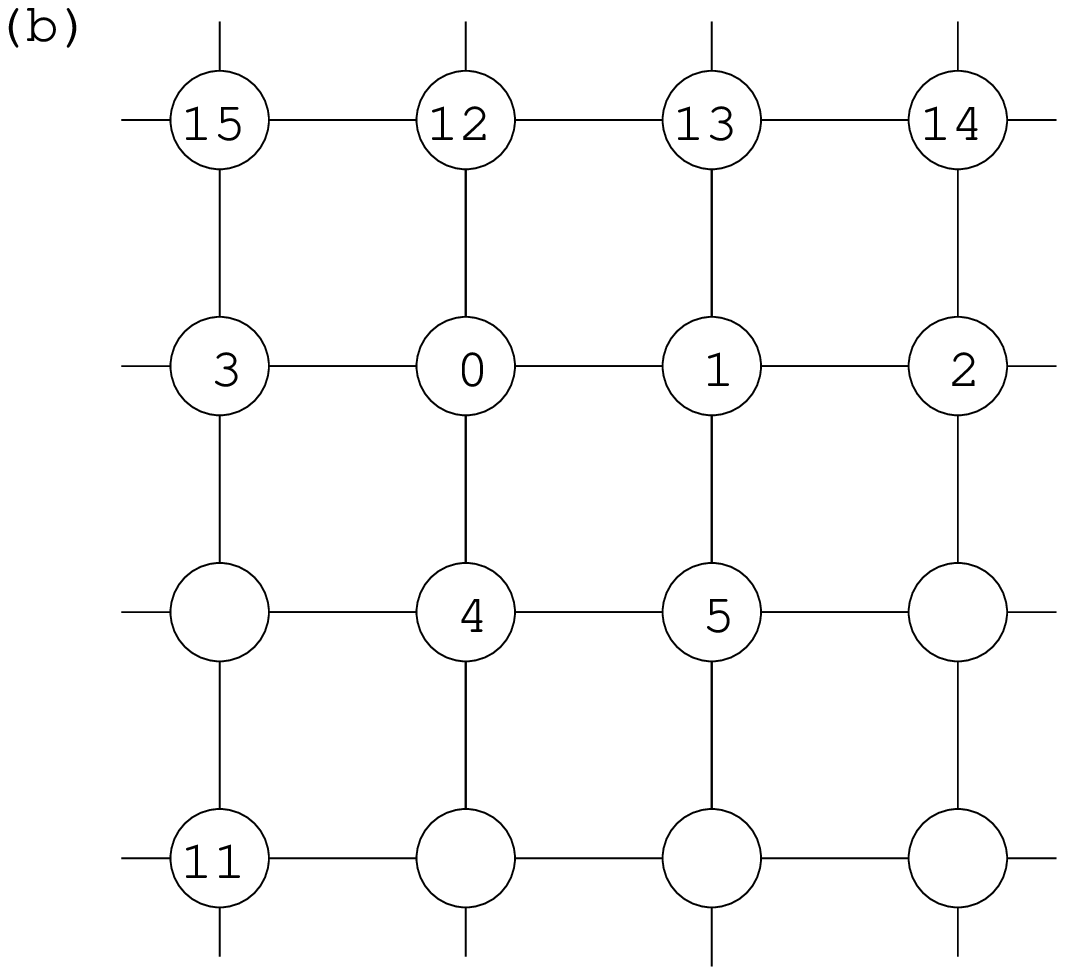}
\caption{(a) A square lattice of size $4 \times 4$ indexed with reference to
the spin at the top left position $(x,y)=(0,0)$. Note that the spin with index
$0$ has {\em two} different shortest paths to the spin with index $2$ due to
the periodic boundary condition.
(b) The indices are translated with a new reference point at $(x,y)=(1,1)$.}
\label{fig:ex}
\end{figure}
For example, let us consider a $4 \times 4$ square lattice with periodic
boundaries, where the spin sites $(x,y)$ can be thus indexed by $k(x,y)=x+4y$,
ranging from $0$ to $15$ [Fig.~\ref{fig:ex}(a)].
In step 1, we may locate $s_i$ at $(x,y)=(0,0)$ on the lattice, so $s_i =
s_{k(0,0)} = s_0$. With respect to this
spin, one can compute $J_{ij}$ for every other spin $s_j$. We exclude
self-interaction by setting $J_{00} = 0$, and $J_{ij}$ should also be zero
when $x=2$ or $y=2$ because their relative displacement is not unique due to
the periodic boundary condition. In this way, one can readily construct the
array $S_0(n)$. Once computed in step 1, it can be used at any spin site
$(x,y)$ to calculate the probability to add another spin to the stack because
the interaction depends only on the relative displacement between them.
the relative position of $s_j$ with respect to
$s_i$ needs to be converted to the actual position on the lattice in step 3(b). Suppose that a
spin at $(1,1)$, i.e., $s_{k(1,1)} = s_5$, is retrieved from the stack.
Then, the situation with this spin as a reference point is equivalent to
Fig.~\ref{fig:ex}(b) under a simple translation. We check which other spins
can be added to the stack by comparing the random number $u$ with $C_j(n)$
in step 3(a) [see Eq.~(\ref{eq:cj})]. If this procedure tells us to add $w =
13$, this correspond to spin $2$ according to the original index in
Fig.~\ref{fig:ex}(a). Therefore, we should add spin $2$ to the stack.
Now, the starting index in Eq.~(\ref{eq:cj}) is changed to $w=13$. Getting
back to step 3(a) and drawing a new random number, let's say that we get
$w=15$ this time. As before, comparing Figs.~\ref{fig:ex}(a) and
\ref{fig:ex}(b), we add spin $0$ to the stack, and go back to step 3(a).
The next $w > 15$ must be beyond the valid range of spin indices, so we stop
considering $s_5$ and retrieve another spin from the stack. This is repeated
until the stack becomes empty.

This algorithm can be extended to simulate $XY$ spins as well:
one should assign a reflection plane by randomly
drawing $\phi \in [0,2\pi)$ on choosing a seed of the cluster, as
originally devised in Ref.~\cite{wolff}. Every spin inside the generated
cluster will be reflected with respect to this plane. We denote this
reflection as an operator $R_\phi$ so that the operation is represented as
$\boldsymbol{s}_i \rightarrow R_\phi \boldsymbol{s}_i$.
Accordingly, whether a spin can be included in the cluster should be also
determined by the energy difference due to such a reflection so that the
added probability becomes $P_{i,j} = 1-\exp \left[ -J_{ij} (R_\phi
\boldsymbol{s}_i - \boldsymbol{s}_i) \cdot \boldsymbol{s}_j /k_B T \right]$.
The LB algorithm for the Ising case first
overestimates $P_{i,j}' = 1-\exp(-2J_{ij}/k_B T)$ and
then adjust it by using the Kronecker delta $\delta_{\boldsymbol{s}_i,
\boldsymbol{s}_j}$. For $XY$ spins, an overestimate occurs in the same
way, but the adjustment should be made by replacing the Kronecker delta with the
probability $P_{\rm add} = \max \left\{ 0, \frac{1-\exp \left[
-J_{ij} ( R_\phi \boldsymbol{s}_i - \boldsymbol{s}_i ) \cdot
\boldsymbol{s}_j /k_B T \right]}{1 - \exp(-2J_{ij}/k_B T)}
\right\}$.

\subsection{Dotsenko-Selke-Talapov Algorithm}
\label{sub:dst}

The DST algorithm was devised to use cluster updates in frustrated systems.
In order to illustrate the main idea, we consider a local version of
Eq.~(\ref{eq:h}):
\begin{equation}
H = J \sum_{\left<ij\right>} \boldsymbol{s}_i \cdot \boldsymbol{s}_j -
3(\boldsymbol{s}_i \cdot \boldsymbol{r}_{ij}) (\boldsymbol{s}_j \cdot
\boldsymbol{r}_{ij}),
\label{eq:loc}
\end{equation}
where the summation runs over all the nearest neighbor pairs~\cite{pra}.
In growing a
cluster $C$, one considers only the first term in Eq.~(\ref{eq:loc}) because it
satisfies $(R_\phi \boldsymbol{s}_i) \cdot (R_\phi \boldsymbol{s}_j) =
\boldsymbol{s}_i \cdot \boldsymbol{s}_j$ for any $\phi$ and does not cause
any energy difference in the bulk of the cluster.
Let us compare two
spin configurations $\mu$ and $\nu$ that are related by one cluster flip, 
i.e., $\boldsymbol{s}_i \rightarrow \boldsymbol{s}_i'$ for every $i$ so that
$\boldsymbol{s}_i' = R_\phi \boldsymbol{s}_i$ for $i \in C$ and 
$\boldsymbol{s}_i' = \boldsymbol{s}_i$ for $i \notin C$.
Then, the ratio of probabilities to select the configurations is
\begin{eqnarray*}
\frac{g(\mu \rightarrow \nu)}{g(\nu \rightarrow \mu)}
&=& \exp \left[ \frac{J}{k_B T} \sum_{\left< ij \right>} (\boldsymbol{s}_i \cdot
\boldsymbol{s}_j - \boldsymbol{s}_i' \cdot \boldsymbol{s}_j') \right]\\
&=& \exp \left[ \frac{J}{k_B T} \sum_{\left< i \in C, j \notin C \right>}
(\boldsymbol{s}_i \cdot \boldsymbol{s}_j - R_\phi \boldsymbol{s}_i \cdot
\boldsymbol{s}_j) \right],
\end{eqnarray*}
just as in the Wolff algorithm. For this generated cluster, we compute an
additional energy contribution from the last anisotropic term,
\begin{eqnarray}
\Delta E^a
&=& \sum_{\left<i,j\right>}
3(\boldsymbol{s}_i \cdot \boldsymbol{r}_{ij}) (\boldsymbol{s}_j \cdot
\boldsymbol{r}_{ij}) -
3(\boldsymbol{s}_i' \cdot \boldsymbol{r}_{ij}) (\boldsymbol{s}_j' \cdot
\boldsymbol{r}_{ij}) \nonumber\\
&=& \sum_{\left<i,j\right> \in C}
3(\boldsymbol{s}_i \cdot \boldsymbol{r}_{ij}) (\boldsymbol{s}_j \cdot
\boldsymbol{r}_{ij}) -
3(R_\phi \boldsymbol{s}_i \cdot \boldsymbol{r}_{ij}) (R_\phi
\boldsymbol{s}_j \cdot \boldsymbol{r}_{ij}) \nonumber\\
&+& \sum_{\left< i\in C, j \notin C\right>}
3(\boldsymbol{s}_i \cdot \boldsymbol{r}_{ij}) (\boldsymbol{s}_j \cdot
\boldsymbol{r}_{ij}) -
3(R_\phi \boldsymbol{s}_i \cdot \boldsymbol{r}_{ij}) (\boldsymbol{s}_j \cdot
\boldsymbol{r}_{ij})
\label{eq:an}
\end{eqnarray}
and accept this cluster move with probability $P_{\rm acc} = \min[1,
\exp(-\Delta E^a /k_B T)]$. Then the acceptance ratios will satisfy
\[ \frac{A(\mu \rightarrow \nu)}{A(\nu \rightarrow \mu)} = \exp( -\Delta E^a
/k_B T), \]
and the transition probabilities in total restore the detailed balance as
\begin{eqnarray*}
\frac{P(\mu \rightarrow \nu)}{P(\nu \rightarrow \mu)} &=&
\frac{g(\mu \rightarrow \nu)}{g(\nu \rightarrow \mu)} 
\frac{A(\mu \rightarrow \nu)}{A(\nu \rightarrow \mu)} \\
&=& \exp \left\{ \frac{J}{k_B T} \sum_{\left< i,j \right>}
\left[ \boldsymbol{s}_i \cdot \boldsymbol{s}_j
- 3(\boldsymbol{s}_i \cdot \boldsymbol{r}_{ij}) (\boldsymbol{s}_j \cdot
\boldsymbol{r}_{ij}) \right] \right\}\\
&\times& \exp \left\{ -\frac{J}{k_B T} \sum_{\left< i,j \right>}
\left[ \boldsymbol{s}_i' \cdot \boldsymbol{s}_j' -
3(\boldsymbol{s}_i' \cdot \boldsymbol{r}_{ij}) (\boldsymbol{s}_j' \cdot
\boldsymbol{r}_{ij}) \right] \right\}.
\end{eqnarray*}

Although this algorithm certainly works, one should note that the cluster
growth does not exactly describe the given system, which means that the
cluster update may not be helpful in overcoming critical slowing
down~\cite{leung,cata}. What usually happens is that a cluster grown to a large
size is simply rejected at the last step, leading to an amount of
inefficiency. Collecting $\Delta E^a$ during the cluster growth may reduce
this problem to some extent~\cite{rosler}: we will check whether this
cluster will be accepted every time $G$ spins are added. Defining
$\Delta E_{ij}^{(1)} = J \left[
3(\boldsymbol{s}_i \cdot \boldsymbol{r}_{ij}) (\boldsymbol{s}_j \cdot
\boldsymbol{r}_{ij}) -
3(R_\phi \boldsymbol{s}_i \cdot \boldsymbol{r}_{ij}) (
\boldsymbol{s}_j \cdot \boldsymbol{r}_{ij}) \right]$ and
$\Delta E_{ij}^{(2)} = J \left[
3(\boldsymbol{s}_i \cdot \boldsymbol{r}_{ij}) (R_\phi \boldsymbol{s}_j \cdot
\boldsymbol{r}_{ij}) -
3(R_\phi \boldsymbol{s}_i \cdot \boldsymbol{r}_{ij}) (R_\phi
\boldsymbol{s}_j \cdot \boldsymbol{r}_{ij}) \right]$, we write down the
cluster algorithm for Eq.~(\ref{eq:loc}) as follows:
\begin{enumerate}
\item Pick randomly a spin and add its index into a stack. Determine a
reflection plane by randomly drawing $\phi \in [0,2\pi)$ and set a variable
$\Delta E_g^a$ as zero.
\item \label{itm:loc} Retrieve an element $i$ from the stack.
\item For every nearest neighbor $j$ of $i$, 
\begin{enumerate}
\item if $j$ is not included in the cluster, 
add it into the stack with probability $P_{i,j} = 1-\exp \left[ -J
(R_\phi \boldsymbol{s}_i - \boldsymbol{s}_i) \cdot \boldsymbol{s}_j /k_B T
\right]$. Add $\Delta E_{ij}^{(1)}$ to $\Delta E_g^a$.
\item Otherwise, add $\Delta E_{ij}^{(2)}$ to $\Delta E_g^a$.
\end{enumerate}
\item If $G$ spins are added into the cluster or the stack is empty,
check whether the cluster can be flipped with probability $\exp( -\Delta
E_g^a/k_B T)$ and then set $\Delta E_g^a$ as zero.
\begin{enumerate}
\item If the answer is no, finish this Monte Carlo step.
\item If the stack is empty and the answer is yes, flip the cluster.
\item Otherwise, go back to step~\ref{itm:loc}.
\end{enumerate}
\end{enumerate}

\subsection{Cluster Algorithm}
\label{sub:com}

We now combine the LB algorithm and the DST algorithm to solve the long-ranged
anisotropic dipole interaction in Eq.~(\ref{eq:h}). We generate a cluster by
using the first term of Eq.~(\ref{eq:h}), which is the same as the LB
algorithm (Sec.~\ref{sub:lb}) except that the interaction becomes
antiferromagnetic due to $J>0$. The DST algorithm (Sec.~\ref{sub:dst}) is
needed to take the remaining terms into account.
Because collecting terms during the cluster growth
takes time of $O(N^2)$ on every retrieval from the stack,
which is highly time-consuming, we
examine the flip after fully generating a cluster at an expense of low
acceptance ratio.
The algorithm can be written down as follows:
\begin{enumerate}
\item Imagine that $\boldsymbol{s}_i$ is placed at the center of the square
lattice, and make an array of partial sums $S_0(n)$ with relative
displacements from this spin site.
\item Pick randomly a spin and add its index into a stack. Determine a
reflection plane by randomly drawing $\phi \in [0,2\pi)$.
\item \label{itm:di} Retrieve an element $i$ from the stack and do the
following:
\begin{enumerate}
\item \label{itm:dcum} Draw a uniform random number $u \in [0,1)$ and find an
index $w$ that satisfies $S_0(w) \le -\frac{1}{2} \ln ( 1-u ) + S_0(z) <
S_0(w+1)$. If there is no such $w$, terminate this loop for $i$. Otherwise,
set $z$ as $w$ for the next iteration.
\item Because $w$ indicates only a relative position with respect to $i$,
translate it into the actual position $w'$.
\item Add $w'$ into the stack with probability
\begin{equation*}
P_{\rm add} = \max \left\{ 0, \frac{1-\exp \left[
J_{ij} ( R_\phi \boldsymbol{s}_i - \boldsymbol{s}_i ) \cdot
\boldsymbol{s}_j /k_B T \right]}{1 - \exp(-2J_{ij}/k_B T)}
\right\}.
\end{equation*}
Go to step~\ref{itm:dcum}.
\end{enumerate}
\item If the stack is not empty, go to step~\ref{itm:di}. Otherwise, go to
the next step.
\item For every spin pair $i, j$ inside the generated cluster $C$,
calculate the energy difference
\[ \Delta E^a_{\rm bulk} = J\sum_{ij} \left[
3(\boldsymbol{s}_i \cdot \boldsymbol{r}_{ij}) (\boldsymbol{s}_j \cdot
\boldsymbol{r}_{ij}) -
3(\boldsymbol{s}_i' \cdot \boldsymbol{r}_{ij}) (\boldsymbol{s}_j' \cdot
\boldsymbol{r}_{ij}) \right] /r_{ij}^3. \]
\item For every spin pair $i \in C$ and $j \notin C$,
calculate the energy difference
\[ \Delta E^a_{\rm surface} = J\sum_{ij} \left[
3(\boldsymbol{s}_i \cdot \boldsymbol{r}_{ij}) (\boldsymbol{s}_j \cdot
\boldsymbol{r}_{ij}) -
3(\boldsymbol{s}_i' \cdot \boldsymbol{r}_{ij}) (\boldsymbol{s}_j \cdot
\boldsymbol{r}_{ij}) \right] /r_{ij}^3. \]
\item Flip the cluster with probability\
\[ P_{\rm acc} = \min \left\{ 1, \exp
\left[ - (\Delta E^a_{\rm bulk} + \Delta E^a_{\rm surface})/k_B T \right]
\right\}. \]
\end{enumerate}
\begin{figure}
\includegraphics[width=0.45\textwidth]{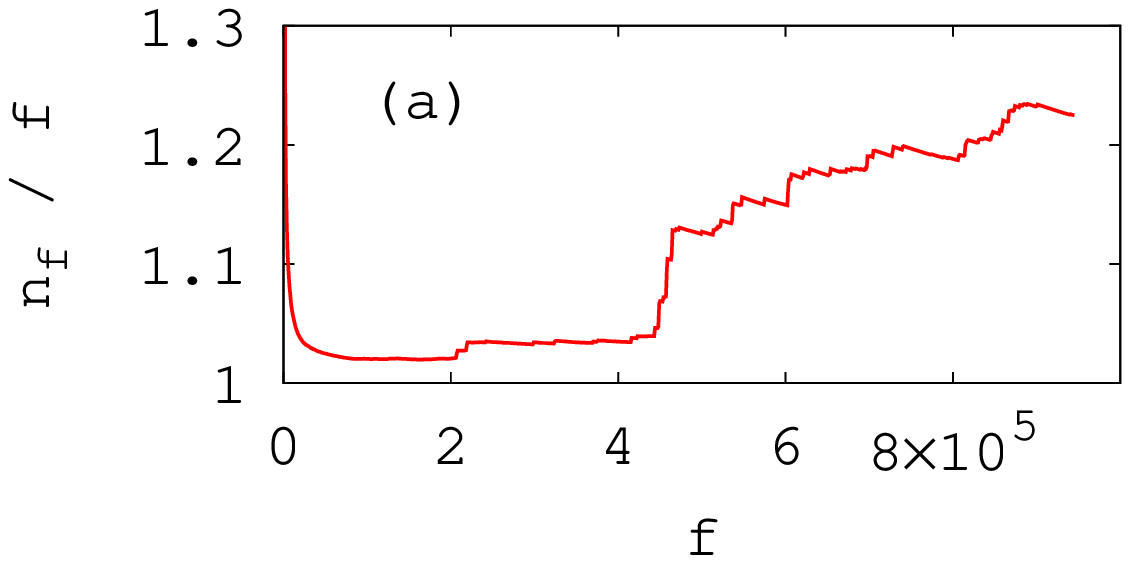}
\includegraphics[width=0.45\textwidth]{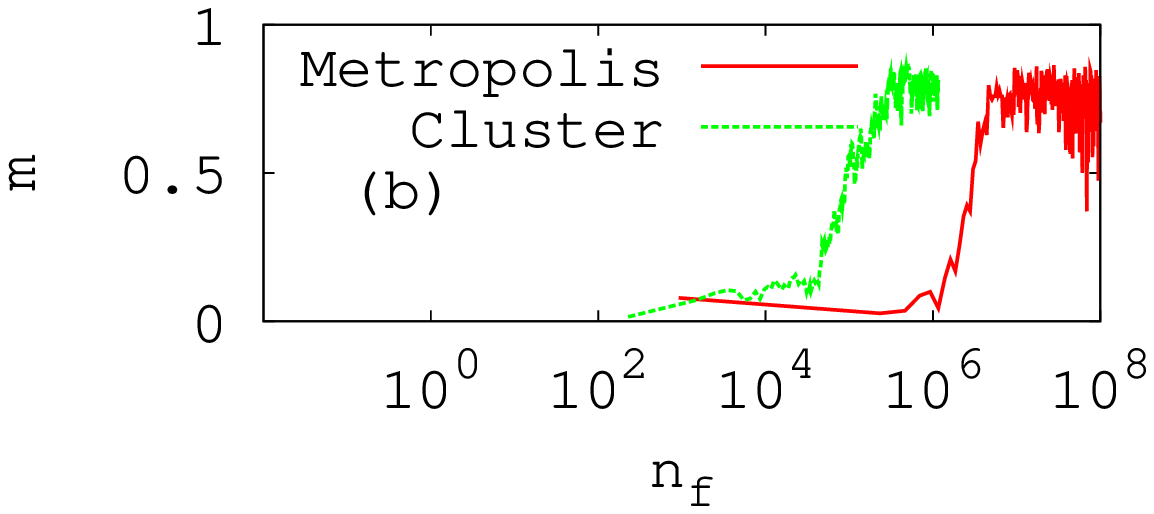}\\
\includegraphics[width=0.45\textwidth]{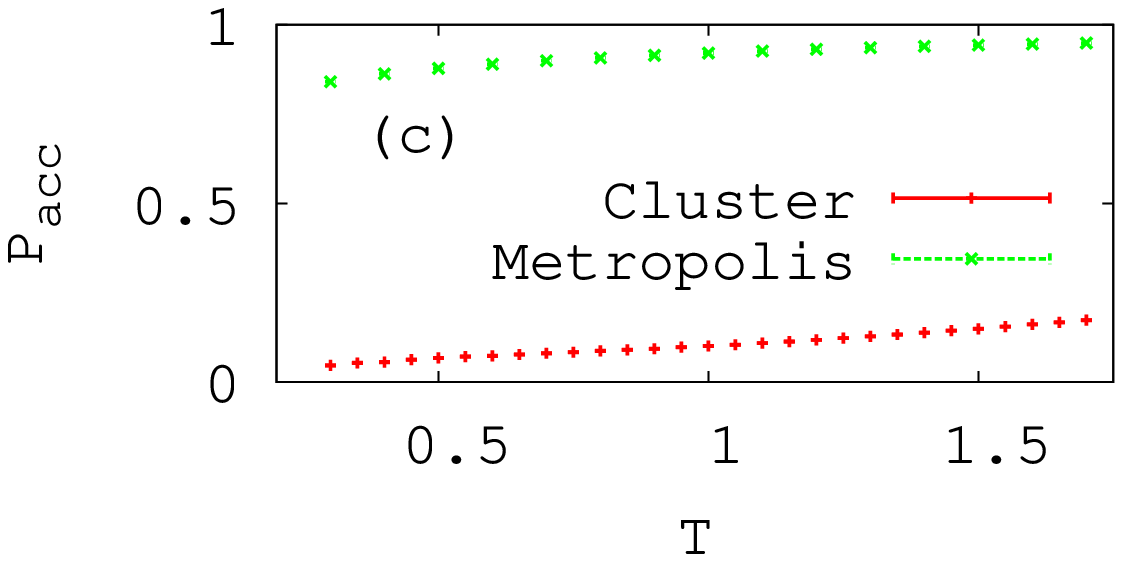}
\includegraphics[width=0.45\textwidth]{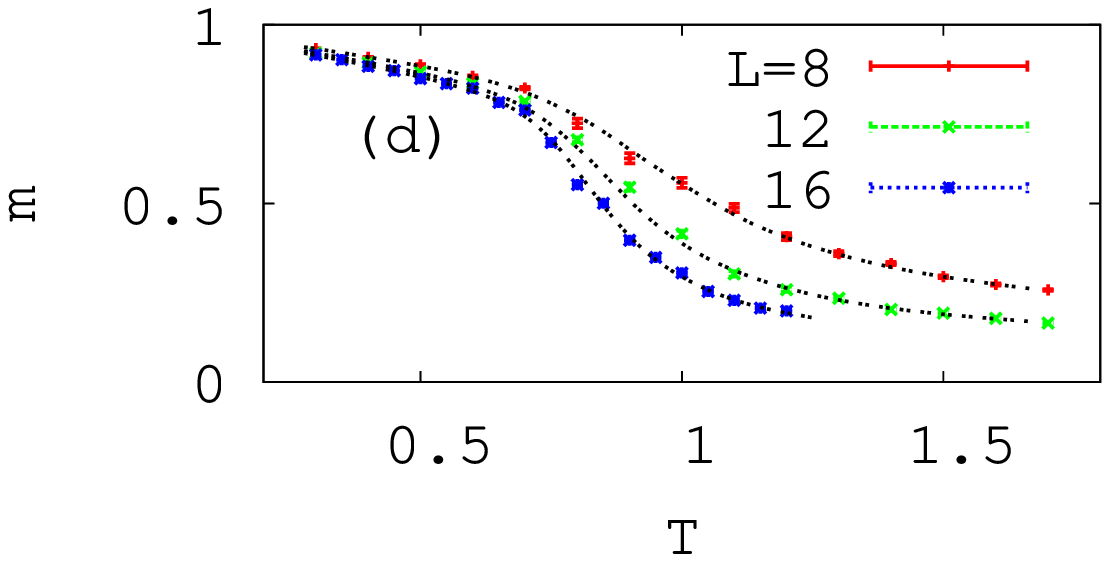}
\caption{The total number of flipped spins is denoted as
$n_f$ and $f$ is the number of cluster flips.
We plot (a) the average number of updated spins per flip and (b) the
magnetic order parameter as a function of $n_f$, measured for $L=16$ and
temperature $T=0.7$ in units of $J/k$.
(c) Acceptance ratios of the cluster algorithm and the Metropolis algorithm.
The system size is taken as $L=8$.
(d) Magnetic order parameter obtained by using the cluster algorithm.
The dotted lines show results based on the Metropolis algorithm for comparison.
}
\label{fig:perf}
\end{figure}

As one sees in Sec.~\ref{sub:dst}, the energy contribution due to
the anisotropy should be calculated inside the cluster and at its surface.
If the generated cluster has a size $c$, the computation for
the bulk part roughly takes $c^2/2$ while the surface part needs $c(N-c)$.
In order for the whole system to be updated, this should be repeated $N/c$
times. Hence, as a whole, it takes $\left[ c^2 + c(N-c) \right] \times N/c = N^2
\left[ 1-\frac{c}{2N} \right]$. In other words, $O(N^2)$ complexity does not
disappear, but decreases to a limited extent. Figure~\ref{fig:perf}(a) shows
how many spins one cluster flip actually updates, which is a small number.
Here, the temperature $T = 0.7 J/k_B$ is chosen to be around the
order-disorder transition point~\cite{honey}.
If we measure time in terms of the number of flipped spins as in
Ref.~\cite{newman}, the magnitude of staggered magnetization,
$m$, is observed to equilibrate substantially faster than the standard
Metropolis algorithm [Fig.~\ref{fig:perf}(b)].
Here, the staggered magnetization is defined as
\begin{equation*}
\boldsymbol{m} = (m_x, m_y) = N^{-1} \sum_i \boldsymbol{\sigma}_i,
\label{eq:m}
\end{equation*}
with
$\boldsymbol{\sigma}_i \equiv \left[
(-1)^{y_i} \cos\theta_i, (-1)^{x_i} \sin\theta_i \right]$,
where the position of each spin is given as $\boldsymbol{r}_i =
(x_i, y_i)$ and the spin variable is written as $\boldsymbol{s}_i =
(\cos\theta_i, \sin\theta_i)$~\cite{debell1}.
We take the magnitude $m = \left| \boldsymbol{m}
\right|$ as the order parameter of this dipole system.
Figure~\ref{fig:perf}(b) implies that the global update indeed carries
out nontrivial moves, even though this factor is largely compensated by the
low acceptance ratio in practical computations [Fig~\ref{fig:perf}(c)].
A trick to get a higher acceptance ratio is to
perform occasionally such a move that rotates every spin in the generated
cluster by $\pi$ because this is the only possible global move that does not
cause $\Delta E_{\rm bulk}^a$. However, this trivial move hardly makes any
essential difference in performance.
Figure~\ref{fig:perf}(d) shows the outcomes from this algorithm, as well as
results based on the Metropolis algorithm for comparison. The nice
agreement found in the order parameter confirms the validity of this
cluster algorithm.

\begin{figure}
\includegraphics[width=0.45\textwidth]{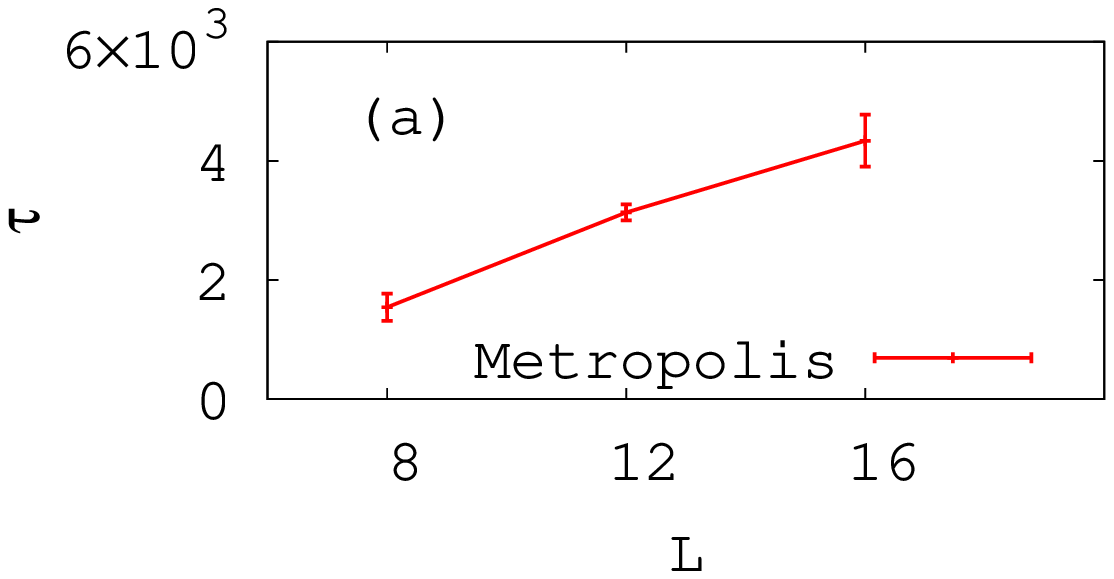}
\includegraphics[width=0.45\textwidth]{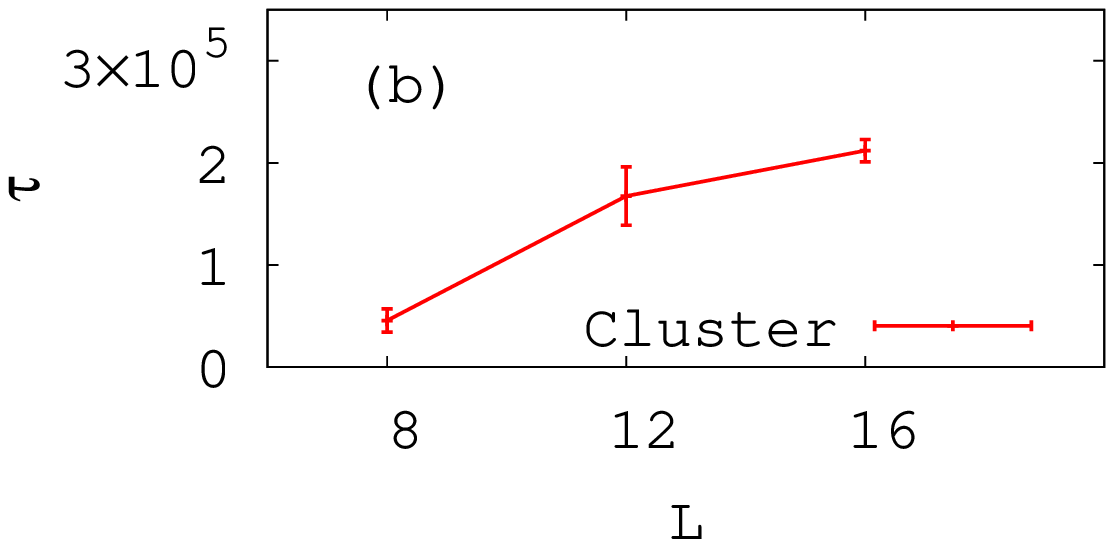}
\caption{(a) Autocorrelation time of $m$ at $T=0.7J/k_B$ in the Metropolis
algorithm, where one Monte Carlo step is defined as attempting to flip every
spin in the system.
(b) The same quantity in the cluster algorithm, where one Monte Carlo
step corresponds to a cluster generation.}
\label{fig:co}
\end{figure}

The autocorrelation time $\tau$ can be measured by integrating the
autocorrelation for an equilibrated time series of $m$. In
Fig.~\ref{fig:co}, we compare $\tau$ of our cluster algorithm with that of
the Metropolis algorithm. We have very limited sizes so it is not easy to
quantify the critical behavior $\tau \sim L^z$. For the Metropolis
algorithm, however, $\tau \approx L$ seems to be a plausible description up
to the sizes used in this work [Fig.~\ref{fig:co}(a)]. On the other hand,
the cluster algorithm shows only a little increase in $\tau$ at $L=16$
[Fig.~\ref{fig:co}(b)], which suggests that $\tau$ can be a sublinear
function of $L$.

\section{DISCUSSION}
\label{sec:sum}

A recent numerical observation based on extensive use of the
Metropolis algorithm suggests that the order-disorder transition of the 2D
square dipole lattice is consistent with the 2D Ising universality
class~\cite{fer}, which has been inconclusive to a large extent.
We have reached the same conclusion by running the Metropolis
algorithm on a number of CPU's in parallel~\cite{honey}. When run on a
single CPU, the agreement of the cluster-algorithm approach found in
Fig.~\ref{fig:perf}(d) is striking, and this shows that the main obstacle
in identifying the critical behavior has been the equilibration rate, as
suggested in Ref.~\cite{tomita}. An efficient algorithm is, therefore, called
for in order to obtain more precise critical properties of the dipole lattices,
and we hope that this cluster algorithm can be a step toward it.

Even though the complexity of $O(N^2)$ still remains in our cluster
algorithm, it has an advantage over the simple Metropolis algorithm when
time is measured by spin flips [Fig~\ref{fig:perf}(b)].
The problem is that it shows little gain in terms of real time due to the
low acceptance ratio, which comes from collecting the anisotropic
contributions. This possibly indicates a direction to improve this
cluster-update approach.

\begin{acknowledgments}
We acknowledge the support from the Swedish
Research Council with Grant No. 621-2008-4449.
This research was conducted using the resources of the High Performance
Computing Center North (HPC2N).
\end{acknowledgments}


\end{document}